# Multimodal concentric surface coils for enhanced sensitivity in MR imaging

Yunkun Zhao[1], Aditya A Bhosale[1], Xiaoliang Zhang[1,2*]
[1]Department of Biomedical Engineering, [2]Department of Electrical Engineering, State University of New York at Buffalo, Buffalo, NY, United States

*Corresponding author:

Xiaoliang Zhang, Ph.D.
Bonner Hall 215E
Department of Biomedical Engineering
State University of New York at Buffalo
Buffalo, NY, 14226
U.S.A.

Email: xzhang89@buffalo.edu

**Abstract**

This study presents the design, simulation, and experimental validation of a novel multimodal concentric surface coil for MR imaging, developed to achieve higher B1 field efficiency while maintaining low SAR for enhanced imaging performance. The coil comprises multiple electromagnetically coupled concentric resonators of varying sizes. The resonant frequency of a desired mode is tuned to 127 MHz, as an example, to demonstrate the performance of the proposed technique at 3 Tesla. Full-wave electromagnetic simulations of the proposed design and bench tests of fabricated prototypes were conducted to evaluate the coil's B1 field efficiency and distribution, multimodal resonance behavior, scattering parameters, and SAR performance. Inductive Current Elimination (ICE) or magnetic-wall decoupling was implemented to enhance channel isolation in a multi-channel configuration to demonstrate the feasibility of applying this multimodal technique to RF array design and parallel imaging. Experimental results show that the proposed concentric coil achieves higher B1 field efficiency and reduced SAR compared to a conventional surface coil of the same size operating at 3 Tesla. Bench measurements on the prototypes confirmed successful tuning and impedance matching, with measured S11 and S21 parameters validating the effectiveness of the decoupling strategy. B1 mapping experiments further demonstrated  efficient RF power delivery across multiple planes. These findings suggest that the proposed multimodal concentric coil has the potential to serve as a promising alternative to conventional surface coils for high-performance MR imaging, offering enhanced RF efficiency, reduced SAR, and the ability to construct multichannel RF arrays.

# 1. Introduction

Magnetic Resonance Imaging (MRI) is a transformative advance in modern medical imaging, offering unparalleled capabilities for visualizing soft tissues and assessing metabolic, functional, and physiological processes in a non-invasive manner [1-5]. Its ability to generate detailed human images without the use of ionizing radiation makes it an essential tool for both investigating fundamental biomedical questions and diagnosing a wide range of medical conditions. MRI has been widely applied across various medical specialties---including neurology, cardiology, oncology, and musculoskeletal imaging---through structural depiction, metabolic quantification, and functional analysis [6-16]. It enables quantitative evaluation of complex anatomical structures such as the brain, heart, and joints [17-27].

Despite continuous advancements, technical challenges persist in achieving high sensitivity or high signal-to-noise ratio (SNR), particularly for imaging at high spatial and temporal resolutions [28-34]. One major limitation lies in improving radiofrequency magnetic (B1) field efficiency, which is linearly proportional to MR detection sensitivity and directly influences image quality and consistency[35-44]. Surface coils, which offer the highest B1 efficiency among coil types (e.g., volume coils [45-47]), often fall short in delivering the required sensitivity for high-quality imaging—especially over large anatomical regions or complex geometries [48-55]. Addressing this limitation is essential for enhancing imaging performance and ensuring reliable diagnostic outcomes [56-58].

Another key challenge is managing the specific absorption rate (SAR), particularly at high and ultrahigh field strengths (3 Tesla and above) [59-63]. SAR quantifies the rate at which RF energy is absorbed by tissues during imaging; excessive absorption can lead to tissue heating and potential safety risks [64-66]. Regulatory SAR limits impose constraints on imaging protocols, often requiring trade-offs between imaging speed, resolution, and patient safety [67, 68]. Since the electric field generated by RF coils is the primary contributor to SAR---with a quadratic relationship to energy absorption---minimizing the electric field to suppress SAR without compromising image quality remains a pressing need [61, 69].

To address these challenges, we propose a novel multimodal concentric surface coil design that leverages electromagnetic coupling among multiple concentric resonators to enhance B1 field efficiency while reducing SAR. This innovative approach integrates multiple resonators of varying sizes, all tuned to the same frequency, into a single flat structure that supports multimodal resonance behavior. One of the resonant modes yields a B1 distribution well-suited for MRI, offering improved B1 efficiency along with reduced SAR and tissue heating.

To validate the proposed design, we conducted comparative studies against conventional surface coils using both numerical simulations and experimental evaluation of fabricated prototypes. The results demonstrate superior B1 field performance, reduced electric fields and SAR, and the feasibility of using the proposed concentric structure as a building block for multichannel RF coil arrays.

## 2. Methods
### 2.1  EM simulation

Figure 1A illustrates the simulation model of the multimodal concentric surface coil and Figure 1B shows the circuit diagram of the concentric coil. The coil design consisted of five electromagnetically coupled concentric resonators, each constructed with copper traces of 6.35 mm width and individually tuned to 194 MHz using dedicated capacitors. The outermost resonator served as the driving coil, connected to a port with an impedance matching circuit. The overall side length of the coil was 9 cm, with individual resonators sized at 9 cm, 7.5 cm, 6 cm, 4.5 cm, and 3 cm. By coupling these resonators, multiple resonant modes were observed, and the mode with the lowest frequency was selected for imaging applications. Through this tuning strategy, the lowest mode of the design was set to 127 MHz, which is optimal for 3T MRI. To evaluate performance, the proposed concentric coil was compared with a conventional surface coil of the same dimensions. This setup allows for direct performance comparisons under identical operational conditions. In simulation, all designs were placed 1 cm below an oil phantom with dimensions of $18 \times 18 \times 9$ cm³ 1cm above all coil designs. The use of the oil phantom allowed for displaying the field efficiency in the region of interest (ROI) under unloaded conditions. Additionally, an alternative simulation comparison was performed using the CST body model Gustav, positioned 1 cm above the coil designs. Performance assessments included analyzing scattering parameters, SAR, and B1 efficiency, with results visualized through field distribution plots. All electromagnetic field plots were normalized to 1 W of total accepted power. Numerical results of the proposed designs were obtained using the electromagnetic simulation software CST Studio Suite (Dassault Systèmes, Paris, France).

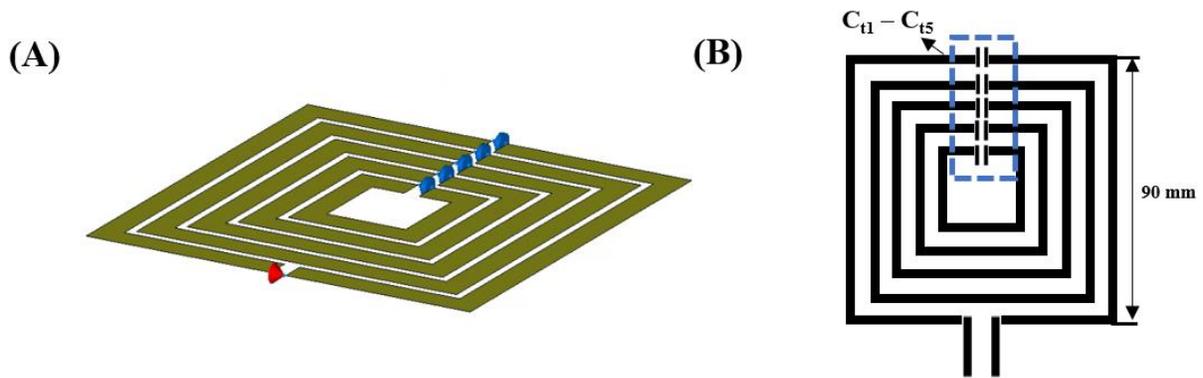

**Figure 1.** (A) Simulated model of the multimodal concentric surface coil. (B) Circuit diagram of the multimodal concentric surface coil.

## 2.2 Model Construction and Bench Test

Figure 2 presents photographs alongside dimensional details of the bench test models for the proposed multimodal concentric surface coil and the comparative designs. These test models were constructed to match the dimensions specified in the simulation models, ensuring consistency between experimental and simulated setups. The multimodal concentric coil was fabricated using 6.35 mm wide copper tape and mounted on a 3D-printed polylactide structure produced with a Flashforge Guider 2s 3D printer (Flashforge, Zhejiang, China). The design was meticulously tuned to a resonant frequency of 127 MHz to align with the operational frequency of 3T MRI systems. Tuning capacitors were incorporated into each resonator to maintain uniformity and optimize performance.

The experimental configuration included an H-field sniffer probe, mounted on a high-precision Genmitsu CNC PROVerXL 4030 router system, to enable precise three-dimensional mapping of the B1 field emitted by the RF coil. The sniffer probe was connected to a Keysight E5061B vector network analyzer (Santa Clara, CA, U.S.), which measured scattering parameters, output power, and accepted power. Data from the network analyzer was transferred to a computer and processed using MATLAB to generate B1 field efficiency maps. Measurements were conducted on a $9 \times 10$ cm slice in the X-Y and Y-Z planes, positioned 1 cm above the coil, as well as a $10 \times 10$ cm slice in the X-Z plane, positioned 3 cm above the coil. The mapping process involved data acquisition at 2.5 mm intervals to ensure high spatial resolution. To standardize results, all measurements were normalized to an input power of 1 watt.

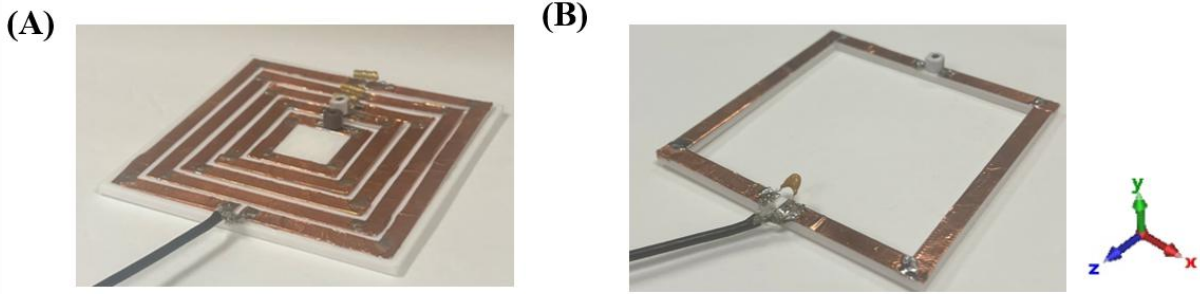

**Figure 2.** Bench test model of the (A) multimodal concentric surface coil and (B) conventional surface coil.

## 2.3 Theoretical Analysis

The proposed coil comprises $N = 5$ identically tuned LC loops of increasing side length, nested coaxially and lying in the same plane. If each isolated loop has self-inductance $L$ and tuning capacitance $C$, the uncoupled resonance is

$$\omega_0 = \frac{1}{\sqrt{LC}}$$

Because the loops are coaxial and carry magnetic flux through one another, every pair ($i,j$) has non-zero mutual inductance $M_{ij}$ ($i \neq j$). We define dimensionless magnetic coupling coefficients

$$k_{ij} = \frac{M_{ij}}{L}, \quad i,j = 1,\ldots,5, i \neq j$$

and collect them in a symmetric coupling matrix $\mathbf{K}$ with unit diagonal and negative off-diagonals:

$$\mathbf{K} = \begin{pmatrix} 1 & -k_{12} & -k_{13} & -k_{14} & -k_{15} \\ -k_{21} & 1 & k_{23} & k_{24} & k_{25} \\ \vdots & \vdots & \vdots & \vdots & \vdots \\ -k_{51} & -k_{52} & -k_{53} & -k_{54} & 1 \end{pmatrix}$$

Writing the complex loop currents as $\{I_i\}$ and assuming sinusoidal steady state, Kirchhoff's laws for the coupled LC network led to:

$$\sum_{j=1}^{5} K_{ij} I_j = \left(1 - \frac{\omega^2}{\omega_0^2}\right) I_i \Rightarrow \left[\mathbf{K} - \left(1 - \frac{\omega^2}{\omega_0^2}\right)\mathbf{I}\right]\mathbf{I} = 0$$

Non-trivial solutions exist when $\lambda$ is an eigenvalue of K with eigenvector I. Each eigenpair ($\lambda_m$,$I^{(m)}$) defines a resonant mode whose frequency is

$$\omega_m = \omega_0\sqrt{1-\lambda_m}$$

The eigenvector $I(m)= [I_1^{(m)},…,I_N^{(m)}]^T$ prescribes the relative phase and amplitude of currents in the loops for mode $m$, which two qualitative facts follow from symmetry: In mode 1 (lowest-frequency), all $I_i^{(1)}$ have the same sign. Physically, currents circulate in the same direction in every ring, producing constructive superposition of their magnetic fields along the coil normal. This yields a loop-like B1 pattern and is the only mode suitable for MR imaging. In higher modes, they contain sign alternation. Some $I_i^{(m)}$ flip sign, creating radial nodes and partial field cancellation, which lowers B1 efficiency and is undesirable for excitation. The central concentration of B1 observed experimentally is also consistent with $I^{(1)}$: smaller inner loops contribute proportionally more near the center, while larger outer loops extend penetration. Small adjustments to relative amplitudes/phases (via capacitor placement and values) can flatten the profile if desired, but the present design intentionally prioritizes high central efficiency.

## 3. Results

### 3.1 Simulated Resonant Frequency and Field Distribution

Figure 3A illustrates the simulated scattering parameters (S11) versus frequency for the proposed multimodal concentric surface coil. The simulation results reveal multiple resonant modes, with the lowest frequency mode occurring at approximately 127 MHz, which is optimal for 3T MRI applications. The additional higher-order modes observed in the scattering parameter plot demonstrate the multimodal nature of the concentric coil design. This tuning strategy ensures that the lowest mode corresponds to the desired imaging frequency, validating the effectiveness of the capacitive tuning applied to each resonator. Additionally, as shown in Figure 3B, the simulated quality factors (Q) of the concentric coil indicate superior performance, with a higher unloaded Q (309.43) and unload/load Q ratio (11.4) compared to the conventional surface coil. This underscores the design's improved efficiency in both unloaded and loaded conditions, highlighting its suitability for 3T MRI applications.

Figure 4A presents the B1 field direction on the Y-Z plane for all resonant modes, where only Mode 1 exhibits a uniform field direction, indicating its suitability for efficient RF transmission. Figure 4B displays the surface current distribution across all five modes, revealing that only in Mode 1 do all five individual coils in the concentric design exhibit current flow in the same direction. These results collectively demonstrate that only Mode 1 achieves the desired field uniformity and consistent current distribution, making it the optimal mode for imaging applications. Figure 5 illustrates the B1 field efficiency for Mode 1 across the X-Y, Y-Z, and X-Z planes. The results confirm that Mode 1 achieves strong and uniform field distribution in all three planes, showing a similar field pattern to the conventional surface coil, further validating its suitability imaging applications.

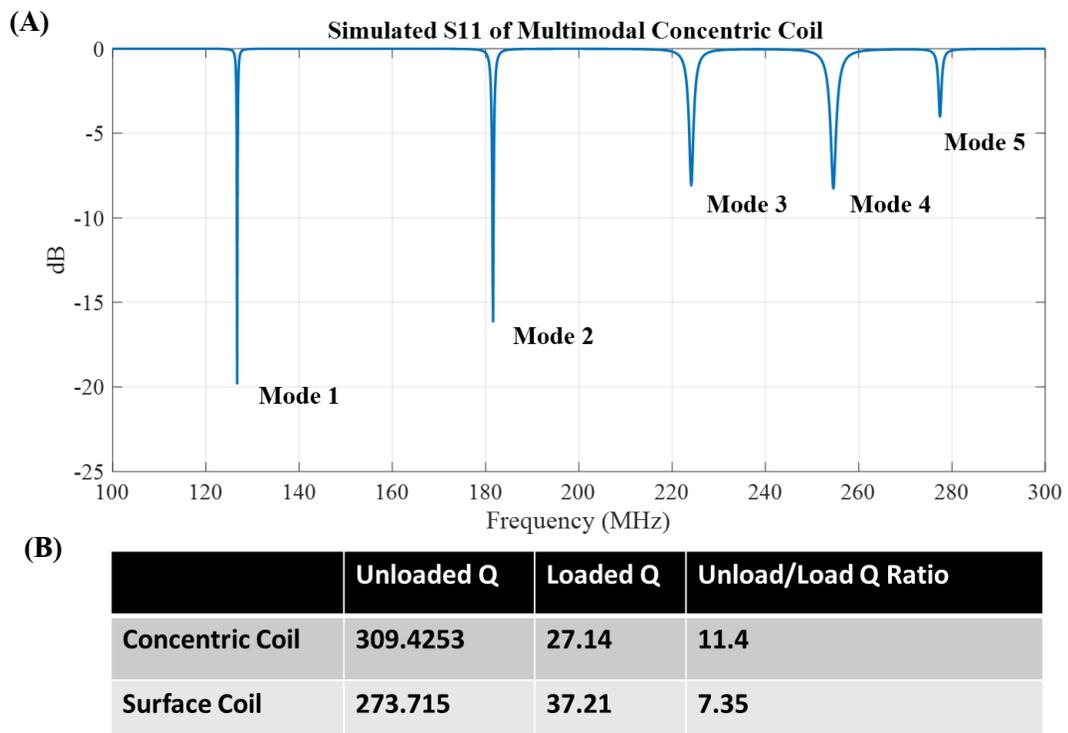

|  | Unloaded Q | Loaded Q | Unload/Load Q Ratio |
|---|---|---|---|
| **Concentric Coil** | 309.4253 | 27.14 | 11.4 |
| **Surface Coil** | 273.715 | 37.21 | 7.35 |

**Figure 3.** (A) Simulated S11 parameters of the multimodal concentric coil (B) Comparison of simulated unloaded and loaded Q factors between the concentric coil and a conventional surface coil

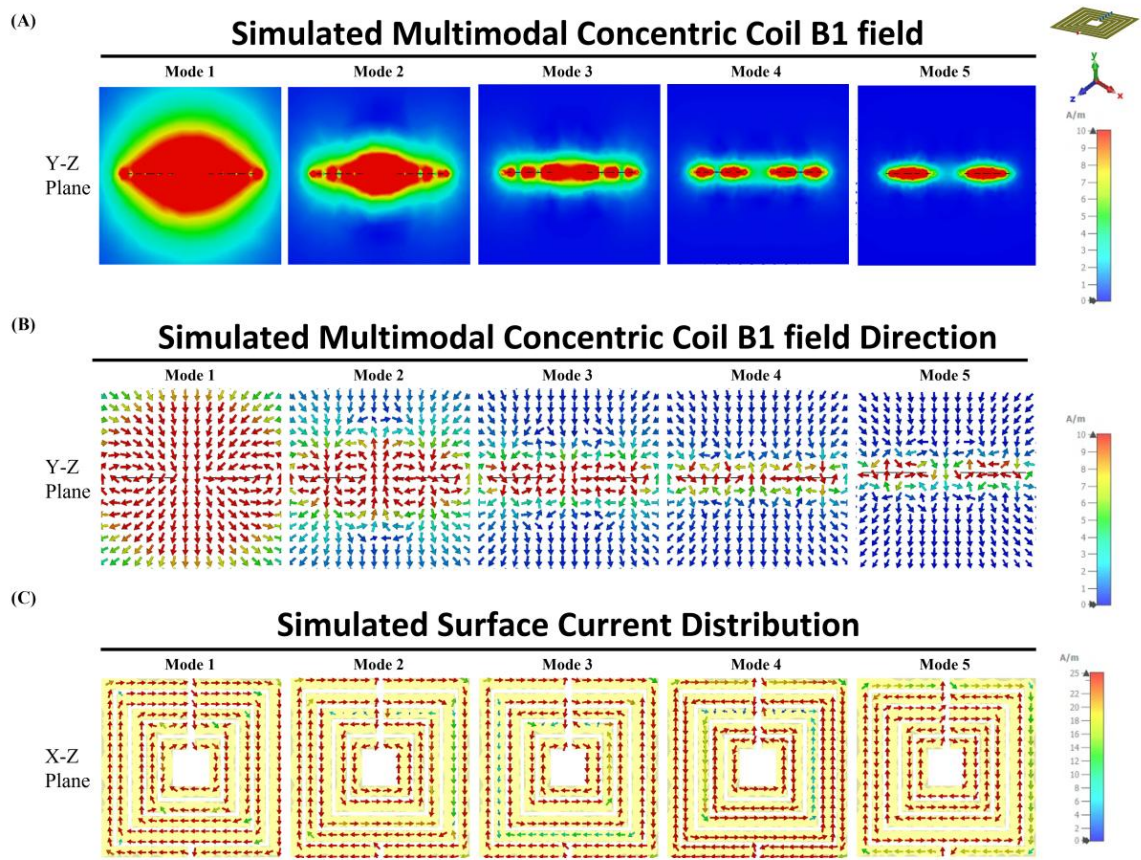

**Figure 4.** (A) Simulated B1 field distribution of the coupled planar array across the Y-Z plane for different resonant modes. (B) Simulated B1 field direction of the coupled planar array across the Y-Z plane for different resonant modes. (C) Simulated surface current distribution across the X-Z plane for each mode.

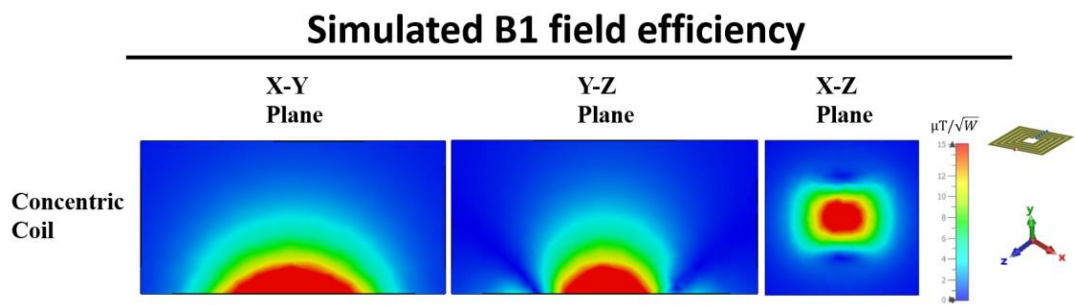

**Figure 5.** Simulated B1 field efficiency of the multimodal concentric surface coil across the X-Y, Y-Z, and X-Z planes.

## 3.2 Measured Scattering Parameters and Field Distribution

Figure 6 presents the measured scattering parameters versus frequency for the multimodal concentric surface coil. The results display multiple resonant peaks, with the lowest peak occurring at approximately 127 MHz, aligning with the simulation findings and the intended operational frequency for 3T MRI. The higher-order resonant peaks observed further validate the multimodal nature of the coil design. These measurements confirm the successful implementation of the tuning strategy, ensuring resonance at the desired frequency.

Figure 7 illustrates the measured B1 field efficiency maps across the X-Y, X-Z, and Y-Z planes, comparing the proposed multimodal concentric surface coil with a conventional surface coil. The efficiency maps show that while both designs achieve a strong B1 field distribution, the concentric coil demonstrates higher B1 field efficiency in all three planes. Measurements were conducted 1 cm above the coils in X-Y and Y-Z plane, and 3 cm above the coil in X-Z plane to assess near-field performance, and the results highlight the ability of the concentric coil to provide superior B1 field efficiency across the imaging area. This increased efficiency further supports the advantages of the proposed coil design in practical experimental setups.

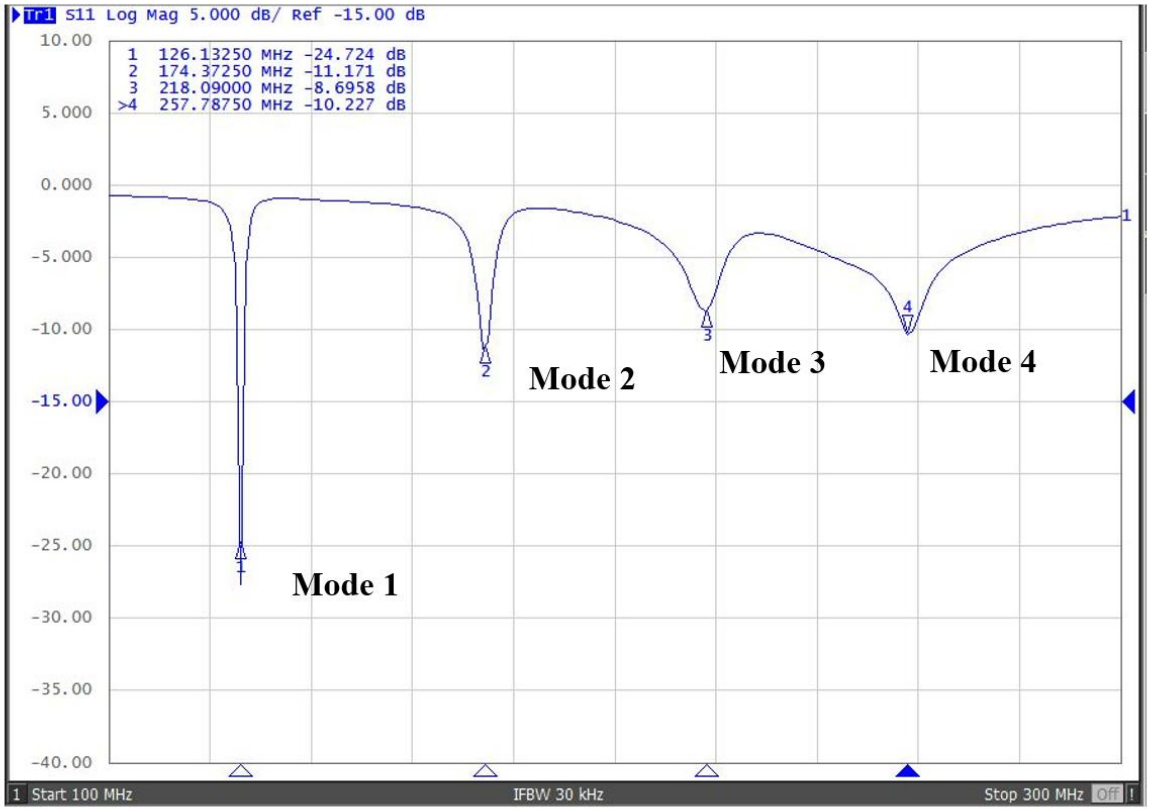

**Figure 6.** S11 reflection measurement vs. frequency of the bench test model of the multimodal concentric surface coil

# Measured B1 field efficiency comparison

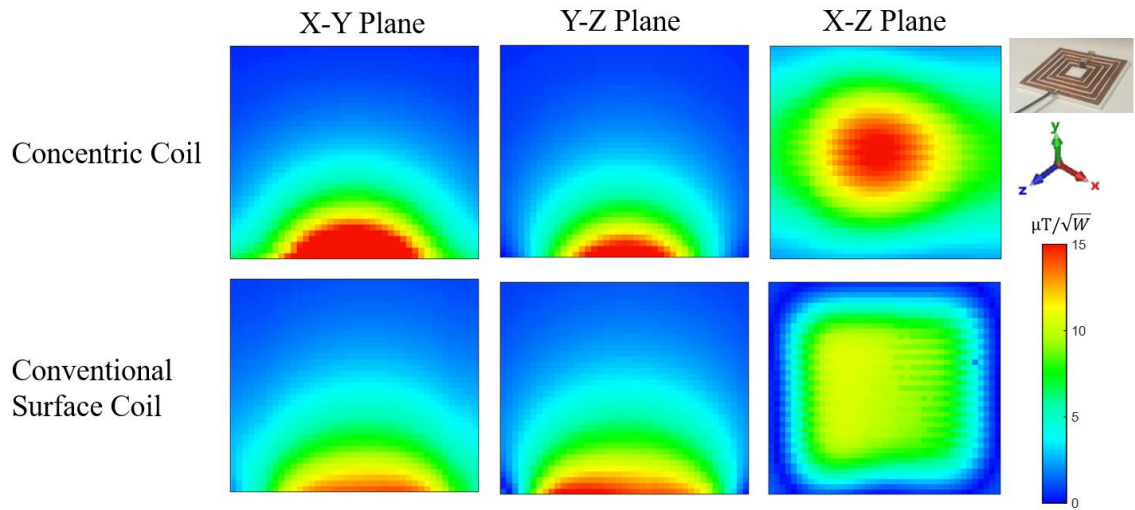

**Figure 7.** Measured B1 field efficiency comparison between the multimodal concentric surface coil and a conventional surface coil across the X-Y, Y-Z, and X-Z planes.

## 3.3 Field Distribution and Efficiency Evaluation

Figure 8A presents a comparison of the simulated B1 field distribution between the multimodal concentric surface coil and the conventional surface coil. The results illustrate that the concentric coil achieves significantly higher B1 field efficiency across the X-Y, Y-Z, and X-Z planes. The field maps indicate a stronger and more centralized B1 field, effectively improving signal strength. Figure 8B shows B1 field efficiency profiles along the X-axis at Y = 3 cm and Y = 5 cm, further illustrating the superior B1 field strength of the concentric coil.

Further evaluation using a human bio-model is shown in Figure 9, where the B1 field efficiency of the concentric coil is compared to that of the conventional surface coil in a realistic anatomical setting. The results demonstrate that the proposed coil maintains superior B1 field efficiency, achieving a stronger B1 field distribution that enhances imaging quality in vivo.

Figure 10 presents the simulated specific absorption rate (SAR) distributions for both coil designs. The concentric coil exhibits a lower maximum SAR of 2.692 W/kg, whereas the conventional surface coil reaches 3.468 W/kg. This reduction in SAR highlights the proposed coil's ability to minimize RF power deposition, improving patient safety while maintaining high imaging performance.

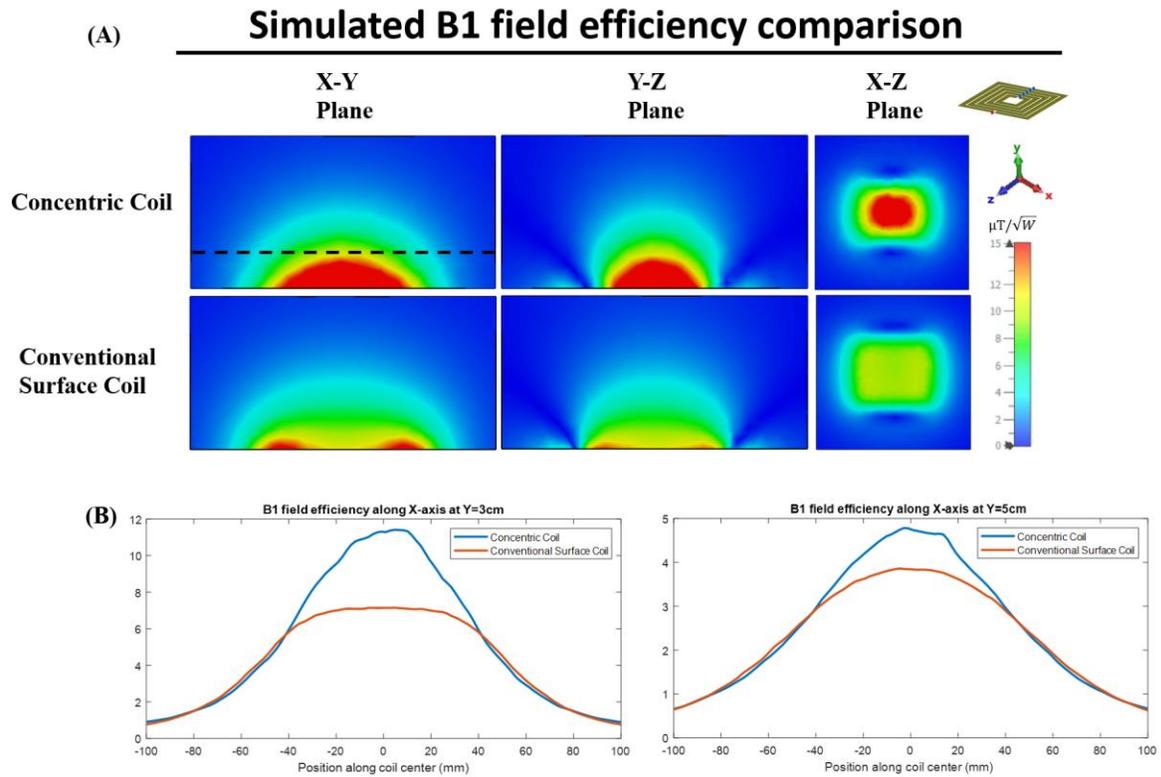

**Figure 8.** Simulated B1 field efficiency comparison between the multimodal concentric surface coil and the conventional surface coil. (A) B1 field distribution across the X-Y, Y-Z, and X-Z planes. (B) B1 field efficiency profiles along the X-axis at Y = 3 cm and Y = 5 cm.

## Simulated B1 field efficiency comparison

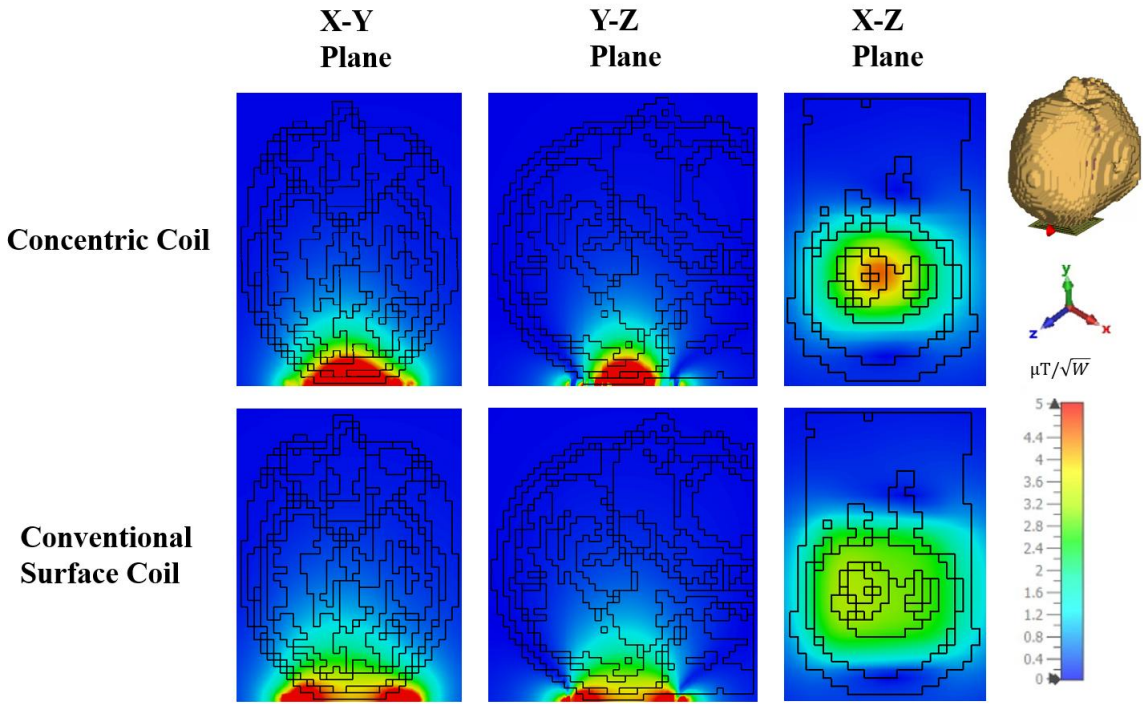

**Figure 9.** Simulated B1 field efficiency comparison between the multimodal concentric surface coil and a conventional surface coil in a human bio-model across the X-Y, Y-Z, and X-Z planes.

## Simulated SAR comparison

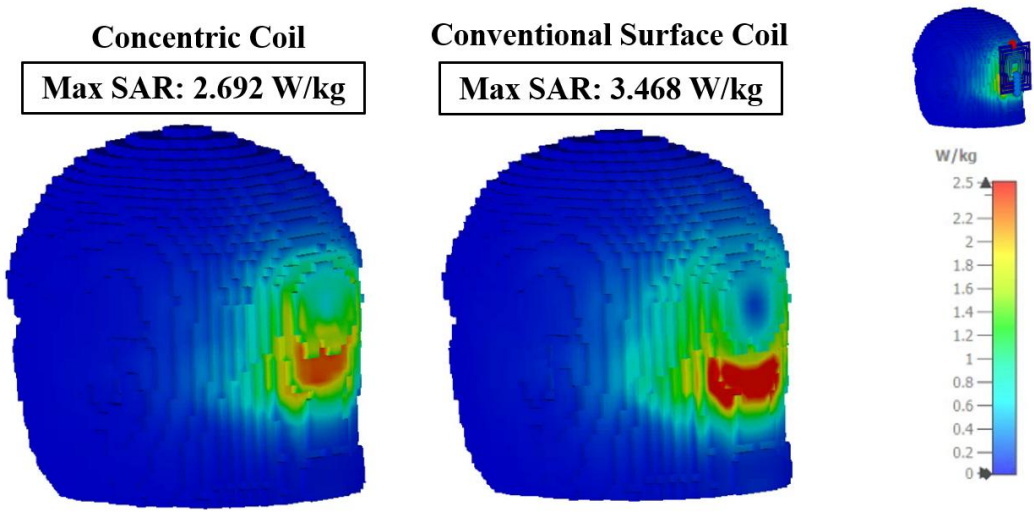

**Figure 10.** Simulated specific absorption rate (SAR) comparison between the multimodal concentric surface coil and a conventional surface coil in a human bio-model.

## 3.4 Decoupling Performance Analysis

Achieving effective decoupling is critical in multi-channel RF coil arrays to minimize channel interference while preserving B1 field efficiency. The proposed concentric coil utilizes Induced Current Elimination (ICE) decoupling [70-77] to enhance channel isolation and improve overall imaging performance. A two-channel bench test was conducted to evaluate the effectiveness of the ICE decoupling strategy, with comparisons of S11 and S21 parameters in configurations with and without decoupling. Figure 11 illustrates the design and setup of the concentric coil integrated with the ICE decoupling circuit used in the bench tests. Figure 11A presents the circuit diagram of the ICE decoupling circuit, while Figure 11B shows the bench test model. The two-channel coil arrangement with ICE decoupling is depicted in Figure 11C. To assess the impact of decoupling, S11, S21, and B1 field efficiency were compared between decoupled and non-decoupled configurations.

Figure 12 displays the bench test results, demonstrating the impact of ICE decoupling on channel isolation and B1 field efficiency. Figures 13A and 13B illustrate the measured S11 and S21 parameters, showing significantly lower S21 values when ICE decoupling is applied, indicating improved isolation between channels. Figure 13C presents the measured B1 field efficiency in the X-Y plane, revealing that ICE decoupling effectively reduces channel interference while maintaining strong and uniform B1 field distribution.

These findings validate the effectiveness of ICE decoupling in multi-channel concentric coil arrays, confirming that it successfully enhances decoupling performance without compromising B1 efficiency. This approach enables improved parallel imaging capabilities and provides a reliable solution for high-field MRI applications.

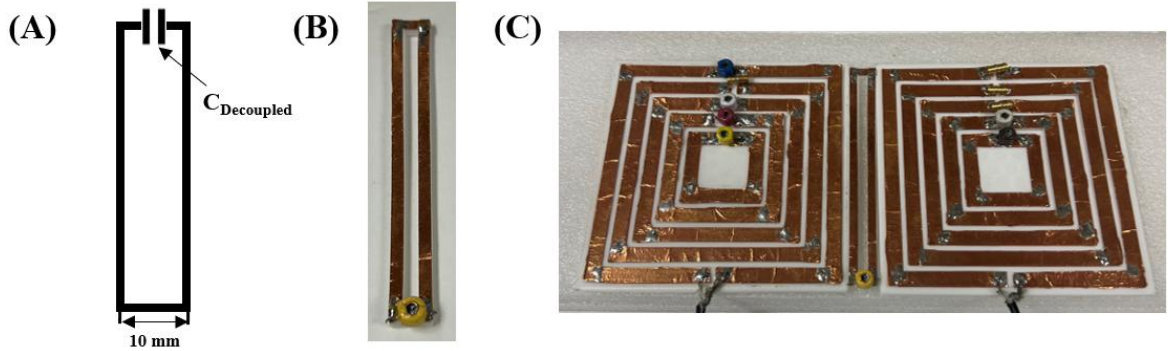

**Figure 11.** Design and setup of the concentric coil integrated with the ICE decoupling circuit used in the bench tests. (A) Circuit diagram of the ICE decoupling circuit. (B) Bench test model of the decoupling circuit. (C) Two-channel concentric coil arrangement with ICE decoupling.

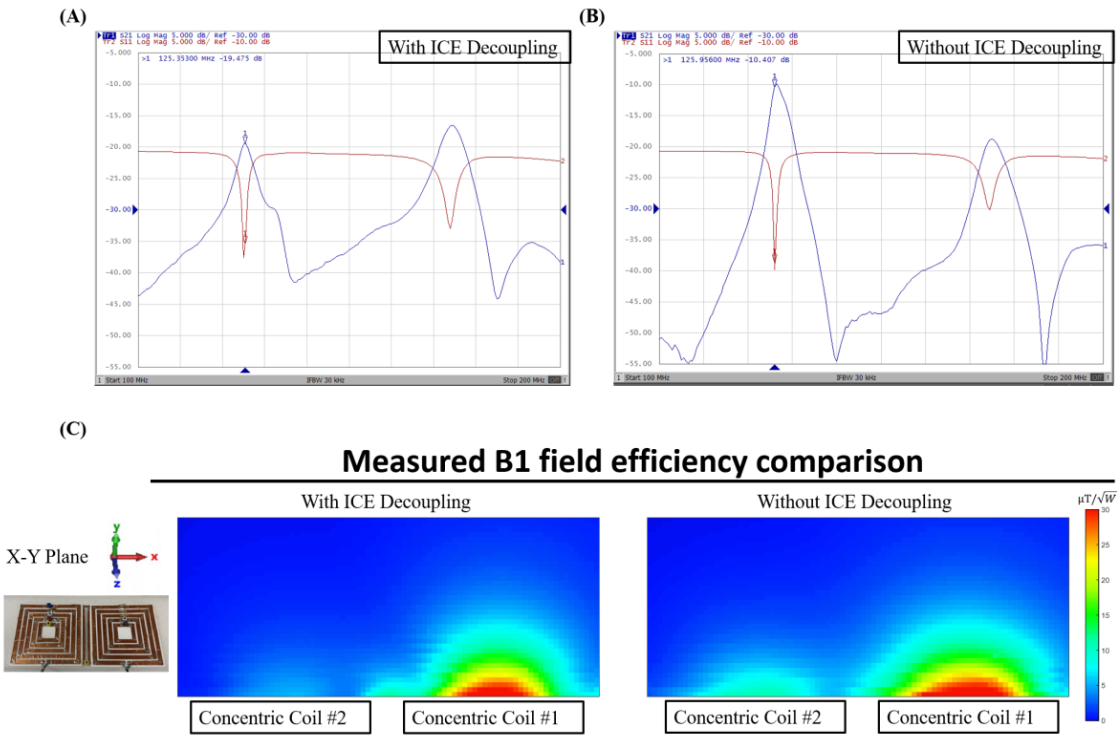

**Figure 12.** Bench test results demonstrating the impact of ICE decoupling on channel isolation and B1 field efficiency. (A) S21 parameters with ICE decoupling applied. (B) S21 parameters

without ICE decoupling. (C) Measured B1 field efficiency comparison in the X-Y plane with and without ICE decoupling

## 4. Discussion

The results of this study demonstrate that the proposed multimodal concentric surface coil offers significant advantages over conventional surface coil designs in terms of B1 field efficiency and SAR performance. The electromagnetic simulations and bench measurements validate the coil's ability to achieve a higher B1 field efficiency while maintaining a lower specific absorption rate (SAR), making it a promising candidate for high-performance 3T MRI applications.

The concentric coil design leverages electromagnetic coupling between multiple resonators, leading to a distinct B1 field distribution. Simulation results show that Mode 1 provides the most effective RF excitation, as confirmed by both B1 field direction and surface current distribution analysis. The measured B1 field maps further reinforce that the concentric coil delivers a stronger and more centralized B1 field, distinguishing it from conventional surface coils that tend to have a more spread-out field. This concentration of RF energy in the center can be advantageous for applications requiring localized high-intensity excitation, though it may require adjustments when uniform coverage is preferred.

Another crucial consideration in RF coil design is balancing field efficiency with SAR constraints. The simulated SAR results indicate that the concentric coil exhibits a lower maximum SAR (2.692 W/kg) compared to the conventional surface coil (3.468 W/kg). This improvement is particularly beneficial for clinical MRI applications, where minimizing SAR is essential for ensuring patient safety and complying with regulatory limits. The ability to enhance RF power efficiency while maintaining lower SAR suggests that the proposed design could enable more flexible imaging protocols and extended scan durations.

While the results are promising, certain limitations should be considered. The centralized B1 field distribution may not be ideal for applications requiring uniform coverage, and future work could explore geometric or capacitive adjustments to tailor the field profile for different imaging needs. Additionally, the current study focuses on single-channel excitation, and further research should evaluate multi-channel driving schemes to optimize flexibility in parallel imaging applications.

# 5. Conclusion

In this work, we presented a novel multimodal concentric surface coil design for MR imaging. One of the multimodal resonance modes exhibits a magnetic field distribution suitable for MRI, closely resembling that of conventional surface coils. The design was validated at a high field strength of 3 Tesla through comprehensive numerical simulations and bench-top measurements, demonstrating improved B1 field efficiency and reduced electric field and SAR compared to conventional surface coils. Our study further suggests that the proposed multimodal concentric surface coils can serve as building blocks for multichannel RF coil arrays when an appropriate electromagnetic decoupling strategy, e.g., ICE or magnetic wall decoupling technique, is applied. By achieving higher B1 efficiency and lower SAR, this approach addresses critical challenges in RF coil design and offers a promising alternative for enhancing MR SNR (and thus resolution), as well as improving patient safety. This work opens new avenues for improving image quality and diagnostic reliability in MRI.

# Acknowledgments

This work is supported in part by the NIH under a BRP grant U01 EB023829 and by the State University of New York (SUNY) under SUNY Empire Innovation Professorship Award.